# Achieve Fully Decentralized End to End Encryption Meeting via Blockchain

Yang, Tan*

Shenzhen Qianhai Xinxin Technology Co., Ltd, t.yang03@mail.scut.edu.cn

Zoom Meeting is an enterprise online video conferencing solution with real-time messaging and content sharing. However, it's lack of privacy protection since centralized Zoom servers are capable of monitoring user's messages. Thereby, to solve the privacy problem, in May 2020, Zoom acquired Keybase so that Keybase's team can help it to build end-to-end encryption meeting while remaining Zoom's current scalability and high-performance. Nonetheless, according to the latest released Zoom's whitepaper, even with the new design of E2E (end to end) encryption meeting, the security threats can't be erased completely since the new design is not fully decentralized.

In this paper, we introduce a fully decentralized design of E2E encryption meeting via blockchain technology. With this new design, Zoom's E2E meeting privacy can be further improved.

**CCS CONCEPTS** • Security and privacy• Security services• Privacy-preserving protocols

**Additional Keywords and Phrases:** Blockchain, Zoom Meeting, End-to-end encryption meeting, Decentralization

## 1 Introduction

Zoom is a well-known company for providing scalable and secure solution for long-distance group video or voice meeting solutions. Also, its product is very easy to use, with a meeting id and password (optional), user can access a meeting by Zoom client, web browser, or phone call on various devices. Zoom has gained a large number of users due to these properties. Especially during the outbreak of the COVID-19 pandemic, more and more people choose to work at home to avoid human contact and Zoom's online group meeting obviously is an ideal solution for them.

However, Zoom's advantages in scalability, easy-use and real-time messaging etc. is built on trusting central Zoom infrastructures which brings a major security issue: central Zoom servers can decrypt user's messages. If a Zoom's server is compromised, a hacker can obtain all meeting's private messages.

To solve the privacy issue, Zoom acquired Keybase in the early of this year. Since Keybase has an experienced team in cryptography and privacy solution, this acquisition marks a key step for Zoom as it attempts to accomplish the creation of a truly private video communications platform that can scale to hundreds of millions of participants, while also having the flexibility to support Zoom's wide variety of uses.

However, according to the description from its latest release whitepaper: "E2E Encryption for Zoom Meetings"[2], even its new E2E encryption design can basically erase the possibility of "Zoom servers can decrypt user's messages", security threats still exist since Zoom doesn't truly achieve a fully decentralized end to end(E2E) encryption. There still are some centralized parties in Zoom's new meeting design. These centralized parties will bring some other security issues which we will cover in the following sections.

On the other hand, blockchain technology emerged in 2008 with the invention of Bitcoin. It's a revolutionary technology that can bring decentralization to the Internet. It solves a long term problem: reaching consensus among different parties in an open, untrusted network. Its security is guaranteed by carefully designed consensus mechanisms along with crypto tools like Hash and Public Key Cryptography. Since decentralized blockchain conforms more to the original design of the Internet, along with the emergence of many other promising projects such as ETH[16], EOS[17], XRP[1], Cosmos[10], etc., it has brought much attention to people and can be applied to various scenarios.

---

* Place the footnote text for the author (if applicable) here.

In this paper, we will propose a truly decentralized E2E encryption meeting protocol via blockchain technology. In our new design, the existence of centralized parties such as SSO IDP, Zoom infrastructures, ZTT auditors are totally removed. Ergo, eliminates the security threats caused by them. The structure of this paper is as follows: First of all, we give an introduction to the background. Secondly, we describe the security and threat model and some basic definitions for our protocol. Thirdly, we will give a brief description of the Zoom's E2E encryption design and the possible security threats it may encounter. Fourthly, we introduce our fully decentralized E2E meeting design and its security analysis.

## 2 BASIC DEFINITIONS AND SECURITY MODEL

In this section, we will give a description of some basic definitions and security model.

### 2.1 Definitions

**PK/SK:** PK: a.k.a Public key for a public key cryptography algorithm and is open for anyone.

SK: a.k.a Secret key or Private key, this key is kept by the user privately.

PK/SK represents a public key pair.

**Meeting participants:** Under the authorization of the meeting leader, the participants who can enter a meeting and access meeting content.

**Meeting leader:** One meeting participant who initiates the meeting and will be considered the meeting leader. This participant will be responsible for generating the shared meeting key, authorizing new meeting participants, kicking out unwanted participants, and distributing keys.

**Adversary:** A malicious user who tries to break the security E2E encryption meeting, he can view all the public information published on blockchain and he may monitor, intercept, and modify network traffic among meeting participants, he may even enter a meeting if he gains legit authorization. However, he does not have direct control over other meeting participants nor access to their secret key. Also, he can't alter the information published on blockchain.

**Blockchain:** Blockchain is a technology to achieve consensus on a single data value or a single state among peers in an open untrusted network.

The name blockchain comes from the fact that the basic data structure of blockchain is made up of a set of blocks. Blocks are ordered by time and chained together through each blocks hash pointer. A block is made up of a set of transactions that happened in a short period.

The most obvious feature of blockchain is decentralization. Without a central party, the blockchain is maintained by a group of peers with equal status. For a public permissionless blockchain like Bitcoin, anyone can join the network and become a peer.

Another obvious character is tamper-resistance. Blockchain can be viewed as an append-only ledger or database. Once a transaction is committed to a blockchain, it can't be tampered.

**Identity Blockchain:** A blockchain that records meeting participants' identity information and mapping relations among meeting participant, device, and device's identity public key.

**Meeting Blockchain:** A blockchain which is used to publish the public information of a meeting which includes: meetingID, all participants' meeting requests, meeting leader, participants' public keys etc.

### 2.2 Security Model

**2.2.1 Certain classes of attack and threats against online meeting.**

There are certain classes of attack and threats against E2E encryption meeting:

**In-meeting impersonation attacks.** A malicious but otherwise authorized meeting participant colluding with other meeting-related parties can masquerade as another authorized meeting participant.

**Metadata and traffic analysis.** Even for end-to-end encrypted meetings, insiders and outsiders can learn details about meeting duration, meeting bandwidth, data streaming patterns, participant lists, and IP addresses.

**Denial of service.** This a very common attack. Attackers can disrupt the message routing and delivering, key distribution etc. to cause the unavailability of the E2E encryption meeting.

In this paper, we will focus on the In-meeting impersonation and Denial of Service attacks and explain how they can be avoided by our new design in the following sections.

### 2.2.2 Security Goals.

Against these adversaries colluding or working independently, an E2E meeting protocol design seeks the following security goals:

**Confidentiality.** Only authorized meeting participants should have access to meeting audio and video streams. People removed from a meeting should have no further access after their expulsions.

**Integrity.** Those who are not allowed into a meeting should have no ability to corrupt the content of a meeting.

**Availability.** The E2E encryption meeting service will always be available for all meeting participants. The adversary can't disrupt the message routing, delivering, key exchange, meeting information publish etc.

# 3 A BRIEF INTRODUCTION OF ZOOM E2E ENCRYPTION MEETING PROTOCOL

## 3.1 Zoom E2E encryption meeting brief description

According to the released whitepaper, Zoom's E2E encryption meeting protocol can be divided into four phases:

**Phase I: Client Key Management.** In the first phase, each user's Zoom device will generate a long-lived PK/SK identity key pair and the central Zoom server will maintain the mapping relation among meeting participants UserID->DeviceID->identity pk. With identity key pair and randomly generated ephemeral PK/SK key pair, users' devices can negotiate and exchange session keys without a trustful central server. Public information needed for the meeting's key negotiation and exchange process is published on a reliable signaling channel which is called "bulletin board" and it is maintained by Zoom servers. Zoom also introduces the meeting leader security codes mechanism to prevent malicious injection of an unwanted public key into this key exchange process.

**Phase II: Identity.** To avoid Zoom server maliciously altering the mapping between meeting participant's UserID to public keys, in this phase, Zoom introduces two parallel mechanisms. Firstly, Zoom introduces SSO and allows SSO IDP to sign a mapping of a Zoom public key to an SSO identity. Zoom can't fake this identity Unless the SSO or the IDP has a flaw. Secondly, Zoom allows users to track contacts keys across meetings.

**Phase III: Transparency Tree.** In the third phase, Zoom implements a mechanism to force Zoom servers (and SSO providers) to sign and immutably store any keys that Zoom claims belong to a specific user, forcing Zoom to provide a consistent reply to all clients about these claims. Each client will periodically audit the keys that are being advertised for their own account and surface new additions to the user. Additionally, auditor systems can routinely verify and sound the alarm on any inconsistencies in their purview. These goals are achieved by building a transparency tree similar to those used in Certificate Transparency[11]and Keybase[8].

**Phase IV: Real-Time Security.** In Phase IV, to avoid malicious Zoom server add a new 'ghost' device for the attacked user who does not have IDP vouch for his identity and fake this user to enter a meeting. Zoom will force a user to sign new devices with existing devices and use an SSO IDP to reinforce device additions, or delegate to his local IT manager. Until one of these conditions is met, Alice will look askance at Bobs new devices.

## 3.2 Potential security threats

From the above description, we can see the new Zoom E2E encryption meeting design has the following security threats:

- First of all, user and his devices' identities authentication heavily relies on the SSO IDP and Zoom which are both central. As it is mentioned in Zoom's whitepaper, if the SSO IDP has a flaw, Zoom can fake user's identity and lead to the in-meeting impersonation attack.
- The append-only property of the ZTT tree which is used to recording mapping relations among meeting participant's UserID->DeviceID->PK is ensured by Zoom and external auditors which are also not fully decentralized. There is a chance that they may collude to modify this mapping relation and lead to the in-meeting impersonation attack.
- Even though Zoom is removed the role of distributing meeting keys but publishing public meeting information on bulletin board and delivering messages among users is still done by Zoom. Those information could be altered by malicious Zoom server and causes denial of service. For example, a malicious Zoom server mixes bulletin board messages from the two meetings and sends previous meeting's ephemeral keys to the meeting leader. In this situation, meeting participants will fail to decrypt the meeting content.
- Even though it has no access to secret key material or unencrypted meeting contents and no longer plays role in distributing initial shared meeting encryption key among participants, Zoom still controls the basic infrastructures to perform the E2E meeting protocols which can cause the unavailability of the service once it is compromised.

To draw a conclusion, because of some centralized roles of the current Zoom E2E encryption meeting design. Security threats such as In-meeting impersonation, Denial of Service still exist.

In the next section, against these security threats, we will proposed a new fully decentralized E2E encryption meeting design based on Blockchain. In this new design, the roles of Zoom, SSO IDP, and ZTT external auditors are removed.

## 4 DECENTRALIZED END TO END ENCRYPTION MEETING

In this section, we will give a description of a new fully decentralized E2E encryption meeting based on Blockchain.

First of all, a simple illustration of the protocol flow can be seen in Fig.1.

In the rest of this section, we will give the detailed description of our new fully decentralized E2E encryption meeting design.

## 4.1 Meeting Participant Identity Registration

(1) Every device of meeting participants will need to generate one unique identity key pair:

$$(IVK_i, ISK_i) = IdentityKeyGen() \quad (1)$$

The algorithm to generate the identity key pair could be any standard, secure public key algorithm, such as Ed25519, Ed448[15], Secp256K1[12]. The public key $IVK_i$ of the identity key pair will be uploaded to the identity Blockchain along with the identity transaction while the secret key $ISK_i$ is kept by meeting participant.·

(2) The identity transaction is of the following form:

$$IdentityTransaction(UserID_i -> DeviceID_i -> IVK_i, UserInfo_i) \quad (2)$$

The mapping information should be attached with a signature

$$Signature = Sign_{ISK_i}(UserID_i -> DeviceID_i -> IVK_i, UserInfo_i) \quad (3)$$

One thing to note, if a meeting participant's userinfo is meant to be private, it can be uploaded in encrypted form by this participant. The decryption key can later be sent to meeting leader or other meeting participants for identity verification via any personal E2Eencrypted channel, e.g., encrypted IM such as Whatsapp, Wechat, Messenger, Signal or Zoom itself.

Furthermore, an alternative option for hiding userinfo can be replacing it by a HMAC value, e.g.:

$$"userinfo": HMAC(r, "Bob, work\ at\ Company\ A, job\ title: Security\ Engineer\ ") \quad (4)$$

This could save more storage for identity Blockchain, and userinfo's plaintext, HMAC key $r$ can be later sent to the meeting leader or other meeting participant for identity verification via any personal E2E encrypted channel.

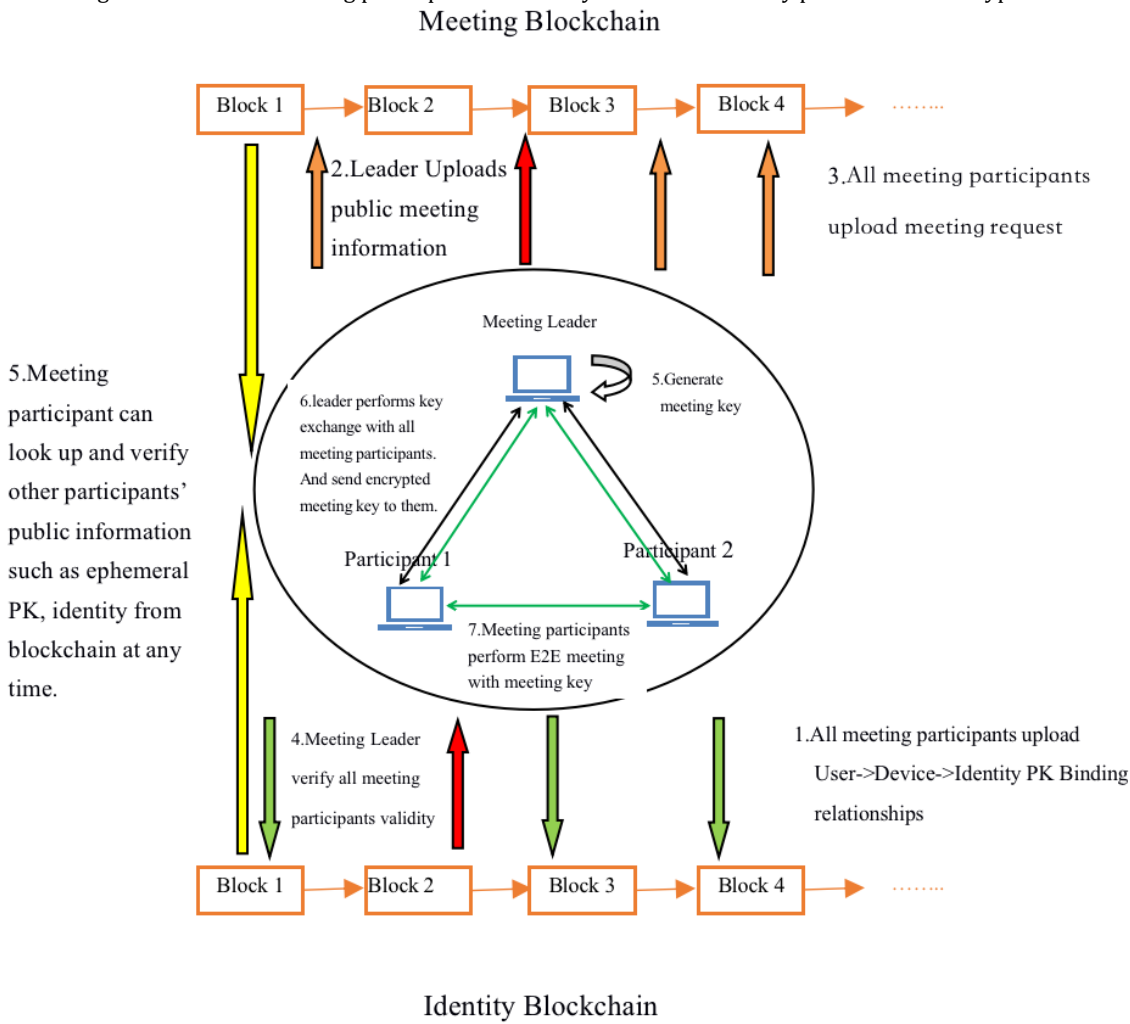

**Fig. 1. Decentralized E2E encryption meeting Protocol Flow**

## 4.2 Meeting Information Publish

If a meeting leader is about to initiate a meeting, he has to publish the public information of the meeting on meeting Blockchain first.

First of all, the meeting leader needs to generate an ephemeral PK/SK key pair for DH key exchange in the "Meeting Key Exchange" process with other meeting participants. The DH algorithm could be any public standard secure algorithms such as Curve 25519, Curve448[13].

$$(pk_l, sk_l) = DHKeyGen() \quad (5)$$

Information which needs to be published includes one randomly generated MeetingID(Unique), Meeting Leader's own identity PK and ephemeral PK, and some other public meeting information MeetingInfo(For example: "this is A company's monthly summary conference").

$$PublishMeeting(MeetingID, MeetingInfo, IVK_l, EPK_l) \quad (6)$$

The meeting information should also be attached with a signature of the meeting leader signed by his identity SK.

$$Signature = Sign_{ISK_i}(MeetingID, MeetingInfo, IVK_l, EPK_l) \quad (7)$$

Once published, everyone who has access to the meeting Blockchain can view the public meeting information, verifying leader's identity. For convenience, the meeting leader can also send this information to the potential meeting participants via any personal E2E encrypted channels.

## 4.3 Meeting Request Generation

Once a potential meeting participant obtains the MeetingID, along with other public meeting information, if he intends to attend the meeting, he has to generate a meeting attending request and upload it to the meeting Blockchain by committing a meeting transaction. The details of this process are as follows:

(1) First of all, after obtaining the Meeting ID, the meeting participant generates an ephemeral PK/SK key pair which later will be used for key exchange.

$$(EPK_i, ESK_i) = DHKeyGen() \quad (8)$$

(2) Each meeting participant generates the Meeting request:

$$MeetingRequest = (MeetingID \mid UserID_i \mid DeviceID_i \mid IVK_i \mid EPK_i) \quad (9)$$

Also, the meeting participant should sign this request with his own identity SK.

$$Sig_i = Sign_{ISK_i}(MeetingRequest) \quad (10)$$

and upload this meeting request along with the signature to the meeting Blockchain by committing a meeting transaction.

## 4.4 Meeting Requests Verifying

In this process, the meeting leader will verify every meeting requests uploaded to the meeting Blockchain.

(1) The meeting leader will obtain all the meeting requests of the meeting he initiated from the meeting Blockchain.

$$GetMeetingRequests(MeetingID) \quad (11)$$

(2) For each meeting participant meeting request, the meeting leader will verify it. First of all, checking the validity of the signature:

$$Verify_{IVK_i}(Sig_i, MeetingRequest) \quad (12)$$

(3) If the signature is valid, then meeting leader extracts the mapping relation

$$UserID_i \mid DeviceID_i \mid IVK_i \quad (13)$$

and verify it from identity Blockchain. If the mapping relation is legit, the meeting leader will decide whether to authorize this meeting participant to enter the meeting by checking this participant's userinfo.

## 4.5 Meeting Key Generation

After verifying all the meeting requests, the meeting leader will generate a 32 bytes meeting key *MK* which later will be used for encrypting the meeting content.

## 4.6 Meeting Key Exchange

In this process, the meeting leader will perform a Diffie-Hellman key exchange[6] with all legitimate meeting participants with his ephemeral key generated in "Meeting Information Publish" process.

The DH key exchange will generate a one-time shared key between meeting leader and each meeting participant which later will be used to send meeting key MK in encryption form.

First of all, meeting leader performs Diffie-Hellman key exchange with every meeting participant:

The meeting leader will compute

$$EncKey_{l_i} = DH(ESK_l, EPK_i) \quad (14)$$

while other meeting participants computes.

$$EncKey_{l_i} = DH(ESK_i, EPK_l) \quad (15)$$

With this shared encryption key, the meeting leader will encrypt the meeting key and send it to every meeting participant.

$$EncMK_{l-i} = Encrypt(MK, EncKey_{l_i})_{aes-256-gcm} \quad (16)$$

When meeting participant *i* receives the encrypted meeting key, he uses the shared encryption key to decrypt and get the meeting key:

$$MK = Decrypt(EncMK_{l-i}, EncKey_{l_i})_{aes-256-gcm} \quad (17)$$

## 4.7 E2E encryption meeting

Once the share meeting key is synchronized to every meeting participant, participants could use this key to encrypt their communication messages. For the forward secrecy, the meeting key won't be directly used for encryption. It will be used to derive the per-stream key by combining a non-secret stream ID using a secure KDF based on HMAC[9]:

$$streamKey = HMAC(MK, streamID) \quad (18)$$

Each streamKey is used to encrypt audio/video UDP packets with AES in GCM mode, with each client emitting one or more uniquely-identified streams.

## 4.8 Meeting key update

During the meeting, if a new meeting participant joins or a participant leaves, the meeting leader needs to generate a new meeting key MK and repeat the "Meeting Key Exchange" procedure to synchronize the new meeting key to every meeting participant.

Similar to the process of "Meeting Request Generation", if a participant leaves, he can generate a MeetingLeave request and upload it to the meeting Blockchain so that the meeting leader and other meeting participants can update the meeting participants list.

One special occasion is when a meeting leader leaves, a new meeting leader needs to be elected. The new leader can be elected by the time order, or manually pointed by the former meeting leader, or by initiating a vote among meeting participants etc. To reassign a new leader, a reassign request should be submitted to the meeting Blockchain.

$$LeaderReassignRequest(MeetingID, IVK_{prev_l}, IVK_{new_l}, EPKnew_l) \quad (19)$$

The request is signed by the new leader's identity secret key.

$$Signature = Sign_{ISK_{new_l}}(MeetingRequest) \qquad (20)$$

If the new leader is assigned by the previous meeting leader, his request will also be attached with the previous leader's signature.

$$Signature = Sign_{ISK_{prev_l}}(MeetingID, IVK_{prev_l}, IVK_{new_l}, EPK_{new_l}) \qquad (21)$$

Once the LeaderReassignRequest is published, other participants will see this request from the meeting Blockchain and check the validity of the request by verifying the signature and checking the agreed leader assigning rules. If the meeting reassign request is validated, the new leader now can perform the "Meeting Key Generation" and "Meeting Key Exchange" procedures and continue the meeting thereafter.

## 4.9 Meeting Dismiss

When the meeting is dismissed or a participant leaves a meeting early, all the meeting related keys should be deleted from a participant's device to ensure the forward secrecy.

In this process, the meeting leader will also generate a MeetingDismiss request and upload it to the meeting Blockchain to inform every other participant that this meeting is ended.

$$MeetingDismiss = (MeetingID) \qquad (22)$$

And the meeting leader should sign this request with his own identity SK.

$$Signature = Sign_{ISK_l}(MeetingDismiss) \qquad (23)$$

To sum up, this is the whole process of the Decentralized E2E encryption meeting based on Blockchain. One thing to note, unlike identity registration, publishing meeting info and meeting requests on Meeting Blockchain can be very frequently which will cause a storage burden for a Blockchain node. To deal with this problem, we can always come up with a solution, for example, Bitcoin's Prune mode. Drop the old meeting data since the meeting data often become obsolete very soon and only keep the recent year's meeting data.

## 5 Security Analysis

## 5.1 How security threats caused by the central parties of Zoom are removed

In this subsection, we will describe how security threats caused by central parties of Zoom are removed by our new fully decentralized design.

Before we get into it, there's a security assumption regarding the Blockchain's tamper-resistance we like to propose.

**Security Assumption** Blockchain can't be tampered with non-negligible probability $\varepsilon$ in time bound $t$, and no adversary can control the majority of nodes of the Blockchain.

Currently, the tamper-resistance of Blockchain is usually achieved by the robustness of Blockchain's Consensus mechanism as well as the scale of the network. To tamper the Blockchain means the attacker can control the majority of the network (51% attack) or reverse a high-level secure hash algorithm like SHA256. In reality, it's only theoretically possible but impractical to own the majority of the network of a large scale of Blockchain such as Bitcoin, Ethereum. Also, it is computationally infeasible to reverse a high-level secure hash algorithm like SHA256[7].

Furthermore, scholars have made some notable progresses on consensus algorithms with strict security proof, such as VDF[3], SSLE[4], we believe they will be put into application soon to achieve true tamper-resistance for Blockchain.

**Theorem** There is no adversary that can successfully perform In-meeting impersonation attack and Denial of Service with non-negligible probability $\varepsilon$ in time bound $t$.

*Proof*
- Unlike Zoom, in our new design, the central Zoom's infrastructure is removed, messaging routing and content delivery can be all done by meeting participants(Blockchain nodes) themselves. If no adversary can control the majority nodes of the Blockchain, the possible denial of service threat mentioned earlier is erased.·
- In our new design, Zoom's bulletin board which is used for publishing public meeting info will be replaced by the meeting Blockchain. According to the security assumption, the meeting Blockchain is tamper-resistant, no adversary can tamper the Blockchain with a non-negligible probability $\varepsilon$ in time bound $t$. Thereby, there will be no malicious Zoom server to modify participant's ephemeral keys. Ergo, eliminate the security threat of denial of service caused by tampering the bulletin board's messages.·
- In our new design, the function of SSO IDP and ZTT transparency tree and ZTT auditors are replaced by the identity Blockchain. Since the identity Blockchain can't be tampered, i.e., no adversary can tamper the identity information on identity Blockchain with a non-negligible probability $\varepsilon$ in time bound $t$. The possibility of ZTT auditors and SSO IDP collude to modify a meeting participant's identity information to perform In-meeting impersonation attack is also removed.

Thereby, the In-meeting impersonation and Denial of Service security threats caused by central parties are removed.

# 6 Possible security challenges in practice and methods to mitigate it

From the previous section, we see how our new E2E meeting design can enhance Zoom's security by introducing the tamper-resistant Blockchain technology.

In this section, despite the security assumption, we talk about a possible security challenge in practice: What if the identity and meeting Blockchain can be tampered. For example, if an adversary successfully perform a 51%[14] attack.

This security threat is especially obvious if these Blockchains are small scale Blockchains. First of all, hackers can easily obtain more than 51% resources of the targeted Blockchain or control the majority of the Blockchain nodes to perform the famous 51% attack. Secondly, without enough developers and community support, bugs of consensus mechanism or implementation errors can also be exploited by hackers more easily.

As to how to mitigate this security threat, the following methods can be considered in practice:

(1) Carefully design the consensus protocol, better with strict security proof such as Algorand[5], VDF, SSLE etc.

(2) Put effort to attract as many users as possible to make it impossible to control more than 51% of the resources of the network.

(3) Anchor its blocks(via block hash) to a large scale Blockchain such as Bitcoin, Ethereum once in a while. This method can directly borrow the security and stability from the underlying Blockchain. A good way to ensure tamper-resistance for a start-up Blockchain.

(4) Build a private chain or consortium chain first, only a few verified nodes have access to the full Blockchain. Other users have to connect to these nodes to obtain information. With more strict access control, it's much easier to isolate malicious users from the Blockchain.

We recommend (3), (4) to build a secure start-up Blockchain in a short time.

# 7 Conclusion

In this paper, we proposed a new fully decentralized E2E encryption meeting design to eliminate the possible security threats for E2E encryption meeting with some central roles, e.g. Zoom.

At a high level, this design is very intuitive, Zoom introduces the E2E encryption design to diminish the role of Zoom's infrastructures in providing online meeting service and enhance Zoom meeting's security. But Zoom's new design can't achieve full decentralization.

Thereby, by introducing the Blockchain technology, our new design can achieve full decentralization and furthermore enhances its security by eliminating the security threats caused by central parties such as Zoom infrastructures, ZTT auditors, SSO IDP, etc.